\documentstyle[aas2pp4]{article}

\def\Msun{\ifmmode M_{\odot} \else $M_{\odot}$\fi}
\def\Lsun{\ifmmode L_{\odot} \else $L_{\odot}$\fi}
\def\eg{{\it e.g.,\ }}

\def\etal{{et al.~}}
\def\etc{{\it etc.\ }}

\lefthead{Chatzichristou}
\righthead{IR-Warm Seyfert Galaxies}

\begin{document}

\title{Multicolour Optical Imaging of IR-Warm Seyfert Galaxies.\\
    V. Morphologies and Interactions. Challenging the Orientation Model}

\author{Eleni T. Chatzichristou}
\affil{Leiden Observatory, P.O. Box 9513, 2300 RA Leiden, The Netherlands}

\affil{NASA/Goddard Space Flight Center, Code 681, Greenbelt, MD 20771}



\begin{abstract}
This paper is the last in a series, investigating the optical properties of a
sample of mid-IR Warm Seyfert galaxies and of a control sample of mid-IR cold
galaxies. In the present paper we parametrize the morphologies and interaction
properties of the host galaxies and combine these with the major conclusions 
in our previous papers. Our results confirm that nuclear activity is linked to
galactic interactions. We suggest an alternative view for the simple 
orientation-obscuration model postulated for Seyfert types 1 and 2, that takes
into account the time evolution of their environmental and morphological 
properties. Within this view, an evolutionary link between starburst-dominated
and AGN-dominated IR emission is also suggested, to account for the 
observational discriminator (mid-IR excess) between our Warm and Cold 
samples.
\end{abstract}




%

\section{Introduction}

Two of the most fundamental outstanding questions regarding the nature of active nucleus galaxies 
are: (i) to what extent can the differences between various observed types of AGN be attributed
to their orientation, evolutionary history or to their intrinsic properties and (ii) what are the
processes responsible for triggering activity in galactic nuclei.
We are exploring these questions on the basis of new observations and analysis of a sample of
IR-luminous Seyfert galaxies, that are sufficiently close that their morphologies, environments
and kinematics can be studied in great detail.
In \cite{paper1} (hereafter Paper I) we have presented imaging data for a sample of Seyfert galaxies
with warm 25-60 $\mu$m colours, selected from the original Warm sample of De Grijp \etal (1987, 1992)
and for a control sample of mid-IR Cold galaxies. In \cite{paper2} (hereafter Paper II) we discussed
their optical properties, resulting from aperture photometry, and their correlations with IR
properties. In Chatzichristou 2000c and 2000d (hereafter Papers III and IV), we parametrized the 
radial light
and colour distributions in the host galaxies.  In the present Paper V, we shall investigate the 
morphologies and close environments of our sample galaxies and will discuss the major results from
Papers II-V, in the context of galactic interactions and mergers. 

A generally accepted unification model 
for AGN postulates that many of their observed characteristics depend upon the orientation of the
observer relative to the dusty torus axis that is surrounding the central black hole
(\eg \cite{antonucci85,antonucci93}).  Our sample is particularly suitable for testing the 
universality of the orientation unification scheme for Seyferts.
Although there have been several previous studies of the morphologies and environments of 
optically-selected Seyfert galaxies, our IRAS-selected Seyferts provide a sample for which dust 
orientation dependent effects should be much less dominant.
Observationally and theoretically there are indications that transient encounters and accretion of 
small companions are responsible for triggering nuclear activity by funneling gas to the central 
black hole (through bar formation for instance) and that violent interactions would rather lead to
complete destruction of the disks involved and trigger large scale star formation events (as the 
ones seen in ULIRGs, \eg \cite{sanders96} and references therein). Optically selected samples of 
AGNs show interactions in only a small fraction of objects. Among IR selected samples this fraction 
is much larger, increasing systematically with IR power (especially when 
10$^{11}\leq L_{FIR}\leq$ 10$^{12}$; \cite{veilleux99}; \cite{gallimore93}). 
Several recent studies leave little doubt that mergers cause IR excess emission,
especially in the most luminous IR galaxies (\cite{sanders99}). Mergers could also be related to 
Seyfert activity, but it remains uncertain what the ultimate fate of the infalling gas will be after
the merger is completed and what are the processes that lead to (re)activation of the central 
engine. 

In this paper we shall give evidence for a causal connection between galactic interactions and
both IR and nuclear activity, suggesting an evolutionary link between starburst dominated and 
AGN dominated IR emission: At the earlier stages of the encounter, material is funneled inwards
triggering intense star formation that produces the bulk of far-IR luminosity. At this stage
the AGN is not yet activated or is hidden by large amounts of dust and the object is 
classified as a Cold galaxy. As gas becomes centrally concentrated the AGN is activated,
rivaling or dominating the luminosity output of the system at mid-IR wavelengths. The BLR
remains still hidden and the galaxy is classified as a Seyfert 2. As time goes by, material is 
consumed in forming stars or blown away by stellar winds; we start seeing the BLR and thus we
classify the galaxy as a Seyfert 1. The occurrence of a mid-IR excess must be also related to
intrinsic properties such as the dust/gas content in the progenitors and the interaction
geometry, than being merely a transition period in the evolution of strongly interacting 
systems. In conclusion, the distinction between Warm Seyfert 1 and 2 galaxies is not one of 
simple orientation. The latter represent an earlier evolutionary stage, with properties
intermediate between the (starburst-dominated) Cold galaxies and the Warm Seyfert 1s and are 
probably affected by (orientation-independent) nuclear obscuration.

Most previous statistical studies of Seyfert environments concentrated on the galaxy density 
within a certain radius, somewhat arbitrarily chosen. As we shall discuss in Section 6 this choice,
as well as the selection of biased control samples, might have serious consequences for the 
results obtained. In the present work we choose an approach that emphasizes the evidence for 
strong interactions between the Seyfert hosts and their neighbours. Since our Warm 
and Cold samples are incomplete and their selection might favour dense environments, we shall
instead (i) probe the environmental differences {\em among} our (sub)samples (ii) investigate 
how these are related to their nuclear- and IR-activity types and (iii) study their 
morphological differences. The present paper is organized as follows:
In Section 2 we describe our parametrization of morphologies and environments.
In Sections 3-5 our results are presented and discussed in connection with the major results
of Papers II-IV. The important points in each section are highlighted.
In Section 6 we compare our results with the recent literature and combine the main 
conclusions of the present study (Papers II-V) in a coherent picture; the implications for the
Seyfert unification models are also discussed. Section 7 summarizes our conclusions.

\section{Parametrization}  

\subsection{\it Environments and Morphologies}

We classify our objects according to their interaction stage and to their morphologies.
For the moment, this classification is phenomenological but as we shall show throughout this 
paper (\eg Section 6) there is a physical basis for this classification. Our interaction
classification (IC) comprises four classes: (I) isolated; (C) objects with companions but no
signs of disruption; (S) strongly interacting systems where at least one of the galaxies is obviously 
distorted with tidal extensions, connecting bridges, \etc ; (Mg) mergers which are (i) either double
nucleus systems or (ii) evolved mergers, where the two nuclei cannot be disentangled but the system 
possesses typical characteristics of a recent merger (for instance tidal tails emanating from the 
main body in opposite directions).
Our morphological classification comprises also four classes: (N) normal morphologies that
is, objects with apparently undisturbed morphologies, no obvious tidal features or bar/ring-like
asymmetries; (B/R) barred/ringed systems (for brevity in some of our graphs we use the label B
although they might have also rings); (T) objects that show some tidal feature (an obviously deformed 
disk, an asymmetric spiral, a one-sided tail etc.) but do not undergo a
disruptive encounter with a close-by companion; (BT) simultaneous occurrence of the last
two classes, which as we shall see is an interesting discriminator by itself. Although the
mergers (interaction class (Mg)) are not part of this morphological classification, they 
are included in our histograms and plots of morphological classes, for comparison with the rest 
of the objects. We allow four types of morphologies for the isolated galaxies and systems with 
companions and two types for the strongly interacting systems.

\cite{mackenty90} classified his sample of Markarian Seyferts in a similar fashion, but his 
definitions are quite different from ours. His interaction classes 2 and 3 are contained in our
interaction class (S) while our class (Mg) is not a distinct class in his classification but 
probably makes part of his interaction class 3 (bridge/tail/jet) or his morphological class 0 
(amorphous); the latter does not correspond to any of our classes.
\cite{whittle92} adopted a similar two-fold classification (see also \cite{dahari85,dahari88}), 
according to the strength of the interaction and the degree of distortion of the host galaxies
and combined the two to a ``perturbation'' class. \cite{hutchings91} considered the interaction
strength and age as the two major parameters in their interaction classification system, following
a more or less subjective evolutionary ``sequence''.
In our approach, the interaction class represents the strength of the {\em actual} tidal perturbation 
between the two galaxies, while the morphological class represents the level of response of the target galaxy
to a current or earlier interaction. Thus, we can recognize an early evolutionary phase (close
encounter and large distortions), as well as, a quiescent phase that presumably corresponds to the end of the merging 
sequence (isolated and symmetric hosts). Any other combination of observed characteristics
is degenerate between interaction strength and interaction age and thus its 
interpretation is subject to an {\em a priori} adopted evolutionary
scheme. In order to disentangle these various factors as much as possible, we have chosen to parametrize
{\em independently} the strength of the gravitational perturbation, as a function of the projected
separation and relative mass of the perturber galaxy (unlike the previous works where these factors
were blended in the interaction classification schemes). This is explained in the following section.

\subsection{\it Interaction Strength}

The classification scheme described above allows for a qualification of the general environmental and 
interaction properties of our samples. We would like to also determine and quantify the factors that
``affect'' the observed nuclear activity type and the development of IR excess in these objects.
We judge whether our target galaxy is isolated or not by
estimating the strength of the tidal interaction with the candidate companion galaxies.
This is proportional to \(\frac{M_s/M_p}{(R/D_p)^3}\),
where $M$ denotes the total mass (disk+halo) and $D$ the major axis diameter of the galaxy and the indices $s$
and $p$ are referring to the companion (secondary) and target (primary) galaxies, respectively;
R is the perigalactic distance (at closer passage) of the perturber galaxy (\eg \cite{byrd86}).
Dahari's 1984 approximation for the galaxy mass was a proportionality with its diameter to the 
1.5 power (appropriate for early type spirals). He thus defined the interaction strength parameter as
\[Q_D=\frac{(D_s\times D_p)^{1.5}}{S^3}\] where $S$ is the projected separation between the two 
galaxies.
As we shall see later, Dahari found a median separation $S$=1.4$D_{p}$ for his (redshift limited) 
Seyfert sample and defined as
``strong'' interactions those with $Q_{D}\geq$1. \cite{byrd86} modeled the gravitational 
instability flows in interacting systems and found that the minimum strength of an interaction 
needed to trigger nuclear activity implies $Q_{D}\approx$0.05 (for low halo systems). Assuming the
minimum encounter separation to be $S_{min}=D_{p}$, Byrd \etal concluded that the minimum dimensions
for a companion galaxy to destabilize the primary's disk is 
\(D_{s(min)}=0.05^{2/3}D_{p}\approx 0.14 D_{p}\) (which, according to these authors, caused Dahari
to miss 50\% of companions for the redshifts of his sample, due to the POSS limitations).  
Assuming an exponential light profile for the companion galaxies, it follows that the limiting 
magnitude is \(m_{s(min)}=m_{p}+4.3 mag\) (for most of the objects we use the $R$ bands). 
For comparison, the interacting galaxies in the Vorontsov-Velyaminov catalogue (\cite{vorontsov59}) 
have a \(\Delta_{m(max)}=m_{s}-m_{p}=3 mag\). 

By using the above criteria, \(D_{s(min)}=0.14 D_{p}\) and \(\Delta_{m(max)}=4.3 mag\), we are more 
likely to include companions that are likely to tidally perturb our target galaxies, than most 
previous studies. We used the conventional 
limiting search radius of 3$\times D_{p}$ for all our objects, which we justify in terms of 
homogeneity, given the limited image sizes. Such a choice is biased towards a particular stage of 
the interaction (close to perigalacticon) and possibly
excludes hyperbolic encounters (the companion can move fast from perigalacticon to large 
distances). However, these will be common limitations for all our samples, the main purpose
of this study being to probe the occurrence and stages of strong interactions in each of them, 
focusing on their inter-comparison.
Since we are not interested in the density of environments but rather in the properties of the
companions that are most likely to have perturbed our target galaxies, we have chosen to
parametrize the brightest companion within our search radius {\em and} the closest companion to 
the target galaxy. In what follows, these will be referred to as the ``brightest'' and 
``closest'' companions. Their magnitudes, dimensions and separations are scaled to those of the
primary galaxy. 
Thus, the relevant parameters in what follows will be: $Q_{D}$, $\Delta m=m_{s}-m_{p}$, 
$D_{s}/D_{p}$ and $S/D_{p}$. For this study we use the totality of our imaged samples that is, 
21 Seyfert 1, 33 Seyfert 2 and 16 Cold galaxies.

\section{Environments and Morphologies}

\begin{figure}
\epsscale{1.}
\plotfiddle{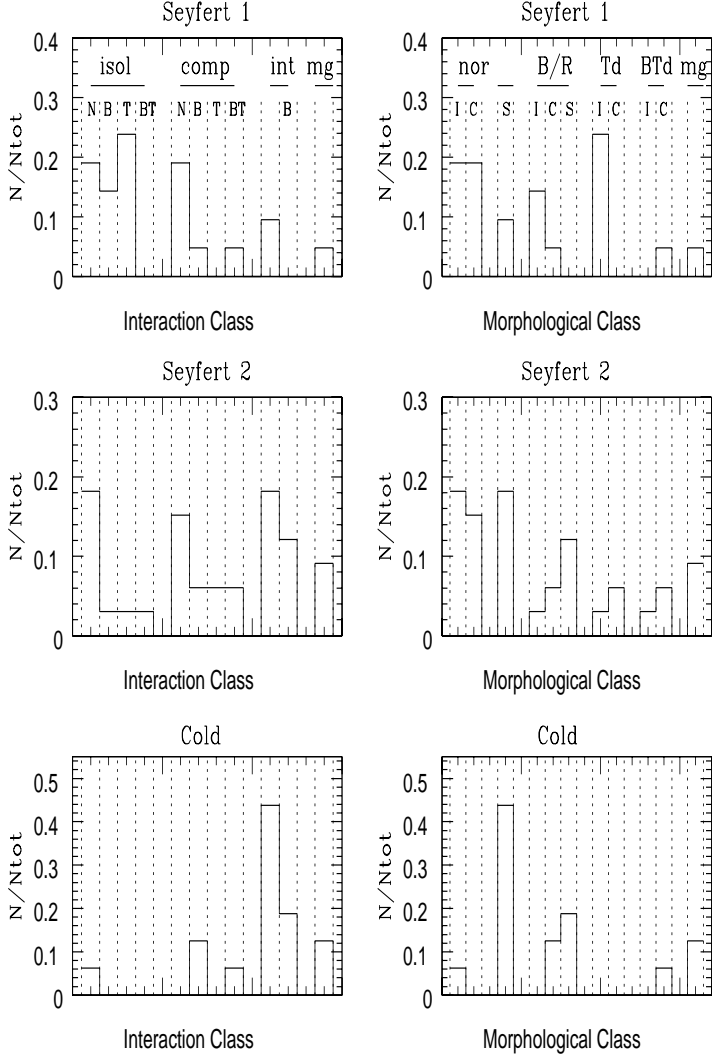}{400pt}{0}{50}{75}{-160}{-120}
\caption{Interaction and morphological classes for our three (sub)samples. We define four interaction classes: isolated (I), companions (C) strongly interacting (S) and mergers (mg) and four morphological classes: normal (N), barred/ringed (B), tidal (T) and barred/ringed plus tidal (BT) features (note that the mergers appear also along the morphological classes). \label{f1}}
\end{figure}

Results are presented in Table~\ref{tab1} and Figure~\ref{f1}.
On the left panels of the latter we show the distribution of the three (sub)samples in interaction 
classes. Within each interaction class we mark the morphological classification. The ordinate 
indicates the fraction of objects within each bin. 
The strongly interacting systems were defined in the previous section, on the basis of their 
environments and distorted morphologies. They are classified in two morphological groups, with or
without bars/rings, as illustrated in Table~\ref{tab2} under morphological classes {\em B/R} and
{\em Normal}, respectively.

\placetable{tab1}

\placefigure{fig1}

The main conclusions of our interaction classification are:

(i) Although there is a relative overlap in interaction properties between the three (sub)samples,
in general Seyfert 1s tend to be isolated, Seyfert 2s include different interaction types and
Cold galaxies tend to be strongly interacting systems.

(ii) 57\% of the Warm Seyfert 1s are isolated and about half of them show tidal features.
From the rest of this subsample, 29\% have companions but do not obviously interact with them
and only 14\% of our Warm Seyfert 1s appear in strongly interacting or merging systems.
Moreover, within each interaction class, bars or rings are not common features.

(iii) Seyfert 2s show a remarkable spread in their interaction properties: approximately
27\% are isolated systems, 33\% have companions, 30\% are strongly interacting and 9\% are
merger systems. Again, within each interaction class, bars or rings are not common.

(iv) Cold galaxies are clearly very different objects or in a different phase in their lifetimes,
compared to the Warm
galaxies: 63\% of them lie in strongly interacting systems and 12\% are mergers. 18\% galaxies
have companions but not disruptive encounters, although all the objects in this class show some
type of instability (bar/ring or tidal) that is likely to be triggered by their companion(s).
Only one Cold object ($\sim$7\%), IRAS 02439-7455, appears in the class of isolated systems.
This is likely to be the result of our definition of companionship, since a smaller galaxy lies
within $\sim$8 kpc projected distance but is 5 magnitudes fainter in $R$, thus was rejected by 
our selection criteria (described in the previous section).
To test the differences between these binned distributions we used the $\chi^{2}$-test, with 
the null hypothesis of matching frequencies between any two samples. We find that the null 
hypothesis can be safely rejected in all cases (significance level $\le$0.01). 

In the same Figure~\ref{f1} (right panels) we show the distribution of the three 
(sub)samples in morphological classes. We find that:

(i) 38\% of the Seyfert 1s show normal morphologies while 19\% are purely barred or/and ringed 
systems. 38\% have some prominent tidal feature, the large majority of which are in isolated
systems. The combination of bar/ring+tidal features is not common (only one object) and 
the same holds for double nucleus merging systems (one object). Seyfert 1s with normal 
morphologies can have any type of environment, while galaxies showing some type of disk
instability are preferentially isolated systems.

(ii) 33\% of the Seyfert 2s appear to have normal morphologies. Among the rest, 9\% (of the 
total subsample) are purely barred/ringed systems and 48\% show at least one prominent tidal 
feature, the majority of which are in strongly interacting systems. The simultaneous occurrence 
of a bar/ring and some tidal feature is more common here (21\% of the total subsample). Finally,
9\% of the Seyfert 2 subsample (3 objects) are double nucleus mergers.  Within each morphological 
class, objects with companions or strongly interacting systems tend to be more common, except
of normal galaxies that can reside in any type of environment.

(iii) Only one Cold object appears to have normal morphology. 69\% of the sample show tidal 
morphologies, among which one third shows also a bar or ring instability and only one object 
does not belong to a strongly interacting system. Two objects (12\% of the sample) are purely 
barred/ringed systems and a similar fraction are double nucleus mergers.

The large spread in morphological properties (causing statistically similar mean absolute and 
standard deviations) for the three (sub)samples, makes it difficult to test the statistical 
significance of the differences between them.
 
In summary, we find a clear progression in environmental properties and interaction stage
between the three (sub)samples: The fraction of isolated systems dramatically decreases from 
Seyfert 1s to Seyfert 2s and to Cold galaxies. The fraction of disruptive encounters
over simple neighbouring is large in Cold galaxies and small in Seyfert 1s, while there is 
almost an equal fraction of them in Seyfert 2s. Similarly, the merger rate increases from Seyfert 
1s through Seyfert 2s to Cold galaxies.   
Thus, it would appear that {\em the distinction between Seyfert 1 and 2 galaxies is not only one 
of simple orientation, but that the latter occur in different environments, ``intermediate'' between
Seyfert 1s and Cold galaxies.} 
This conclusion is further reinforced by the comparison of host morphologies: There is a 
dramatic decrease in the fraction of normal host morphologies between the Warm and Cold samples, 
although this is not very different between the two Seyfert types. The fraction of 
hosts showing tidal disturbances increases from Seyfert 1s through Seyfert 2s to 
Cold galaxies; since in the former case most of these galaxies are isolated while in the latter
are strongly interacting systems, this means that also the {\em shape} of tidal features is 
different in Seyfert 1 hosts (one-sided tails or fans) compared to Seyfert 2 and Cold hosts 
(tidal arms, bridges and tails).
Finally, if the presence of bars and rings is independent of environmental properties or 
interaction stage, there is a progressively larger fraction of these features from Seyfert 1 
through Seyfert 2 to Cold galaxy hosts. However, if we exclude the strongly interacting systems,
this fraction is similar for Seyfert 1s and 2s . The latter observation and the fact that
similar fractions of Seyfert 1s and 2s have normal hosts in our samples, may indicate that
{\em the orientation unification applies for some of the Seyfert 2 galaxies, but a fraction of 
these (the strongly interacting) are intrinsically different objects than the Seyfert 1s.}

\section{Interactions vs Other Properties}

To further investigate the Seyfert type 1 vs 2 relation, we searched for connections between
environments and host morphologies and the optical and IR properties, for our three (sub)samples. 
In Figure~\ref{f2} we plot some of the most important correlations.

\placefigure{f2}

\subsection{Optical Properties}

(i) There is a significant correlation between Seyfert 2 optical (in particular disk) luminosities
and interaction class (top row left panel). Moreover, within each interaction class, 
bars/rings or tidal features further increase the disk luminosity (top row right panel). 
{\em In Paper II we had found that Seyfert 2 disks are larger and brighter than Seyfert 1 disks;
this seems to be related to the largest fraction of interacting systems among the former.} 

(ii) Seyfert 1s, on the other hand, show
no correlation of nuclear optical luminosities with interaction class, but tidal and 
barred/ringed morphologies seem to increase their {\em disk} brightness (top row right panel). 
However, this is small number statistics since we have photometric measures for few non-isolated 
Seyfert 1s. Indeed, when we use the integrated $V$-band magnitudes given by \cite{grijp92},
we find a tendency for Seyfert 1s to become optically {\em fainter} at more advanced interaction 
class (second row from top, left panel). This is most likely an effect of increased dust 
extinction in these systems.

(iii) Nuclear and disk colours are uniformly distributed for Seyfert 2s in all 
interaction/morphological classes (third row from top, left panel). For Seyfert 1s however, their 
nuclei are 
significantly bluer for isolated galaxies, compared to the other interaction classes for which
colours are similar to the Seyfert 2s. {\em In Paper II we have shown that Seyfert 1 nuclei tend to 
be overall bluer compared to Seyfert 2s, which we demonstrate here is related to the larger 
occurrence of the former in isolated galaxies.}

(iv) Seyfert 2s show a clear trend for their inner colour gradients to become bluer with 
interaction ``strength'' (third row from top, right panel). There are too few Seyfert 1s with 
measured gradients to assess any such correlation. {\em As we have pointed out in Paper IV, 
Seyfert 2 colour gradients are indicative of strong starbursts, that we demonstrate here to be 
related to galactic interactions.} 

(v)  For the Cold sample, there is no clear trend for optical luminosities to correlate with 
environment or morphology, but they are on the average fainter than Seyfert 2s (and Seyfert 1s)
for similar interaction/morphological classes (top row panels). 

(vi) In Papers II and III we have seen that Seyfert 1s are more centrally concentrated compared to
Seyfert 2s and Cold galaxies. In Figure~\ref{f2} (second row from top, right panel) we see 
that the degree of light concentration is also a function of morphological class, this being a 
stronger effect for Seyfert 1s. 

(vii) We do not observe any
bimodality in any of the above properties or correlations in Seyfert 2s but rather continuous
distributions that lie between the (generally) different loci of Seyfert 1 and Cold galaxies
in most of these diagrams.  

\subsection{IR Properties}

\begin{figure}
\epsscale{1.}
\plotfiddle{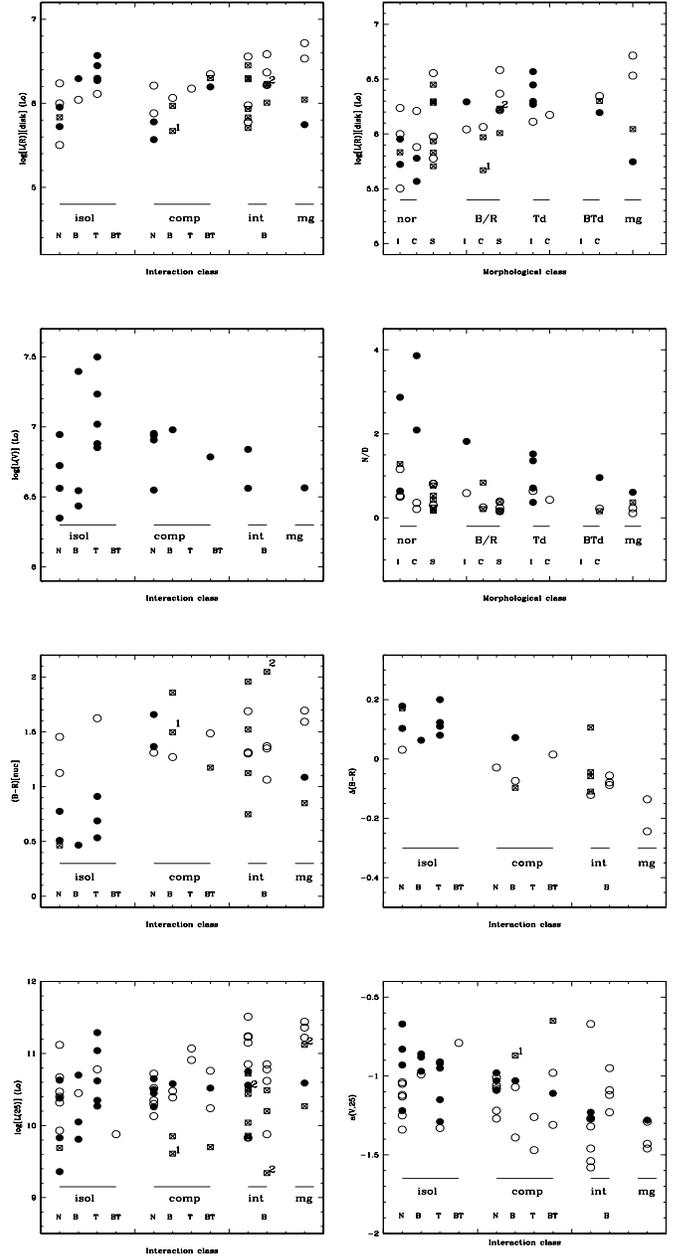}{400pt}{0}{63}{63}{-190}{-30}
\caption{Optical and IR properties versus interaction and morphological classes. Filled circles are Warm Seyfert 1s, open circles Warm Seyfert 2s and crossed squares are Cold galaxies. (The left panel in the second row from the top shows integrated $V$-band luminosities from De Grijp \etal 1992 for the Seyfert 1 subsample (see text)). \label{f2}}
\end{figure}

(i) We find that the 25 $\mu$m emission is correlated with interaction class for the Warm Seyfert 
2s and possibly for the Cold galaxies too (Figure~\ref{f2}, bottom left panel). A similar good 
correlation of $L_{FIR}$ with interaction class exists for the Cold sample (not plotted). The latter 
is not obvious for the Seyfert 2s, except for the merger systems that have consistently higher IR 
luminosities at all wavelengths, compared to the rest of the sample. For the correlation shown in 
Figure~\ref{f2}, within each interaction class the objects showing tidal features have 
larger $L_{25}$ compared to the barred/ringed and normal morphologies, this being also true for
the Warm Seyfert 1 subsample.  

(ii) In isolated Seyferts or those with normal morphologies, the $L_{25}$ is comparable for the 
two nuclear types but for tidal morphologies, strongly interacting or merging systems, $L_{25}$ is 
significantly enhanced in Seyfert 2s (bottom left panel). This behaviour would be consistent with
a bimodality within the Seyfert 2 subsample. {\em In Paper II we had found that Seyfert 2s show 
overall stronger IR emission compared to Seyfert 1s and we had attributed this to warm dust in 
their disks. Here we show that this result is probably due to the larger fraction of strongly 
interacting systems within the Warm Seyfert 2 subsample.}

(iii) Cold galaxies have lower $L_{25}$ than both Seyfert 1s and 2s, for the same interaction or 
morphological class, while in general similar (or larger) $L_{FIR}$ luminosity.

The most straightforward interpretation for the above correlations is that {\em the 25 $\mu$m 
emission in Seyfert galaxies is mainly due to warm dust heated by the AGN, unless these are
strongly interacting systems in which case disk star formation causes significant excess 
25 $\mu$m emission in Seyfert 2s. Far-IR emission on the other hand, is mainly due to warm dust in 
the host disks and is similarly enhanced by strong interactions in Warm and Cold galaxies.}
This would be consistent with the observed absence of a correlation between 25-60 colour indices 
and interaction or morphological class: {\em Although the frequency of disruptive encounters is an
important discriminator between samples selected according to their mid-IR colours, the strength 
of these encounters is not the only factor determining the mid-IR excess}. 

(iv) There is an important correlation found between IR loudness and interaction class for Seyfert 1s,
shown in Figure~\ref{f2} (bottom right panel). In Paper II we have defined the IR 
loudness using both the nuclear and total optical $V$ magnitudes. Here we find that Seyfert 2 (and 
Cold) {\em nuclei} are IR louder compared to Seyfert 1 nuclei independently of interaction class, 
which supports the suggestion put forward in Paper II, that the {\em nuclear $\alpha_{(V,25)}$ is 
primarily a measure of dust extinction in Warm Seyfert 2s}. Isolated Seyfert 2 galaxies are 
also IR louder than Seyfert 1s when considering the {\em total} $\alpha_{(V,25)}$ but for the 
other interaction classes there is a larger scatter of values since the dust obscuration and 
25 $\mu$m enhancement both affect this index in Seyfert 2s. In non-isolated Seyfert 1s 
on the other hand, IR-loudness correlates tightly with interaction class, this being mainly due
to the decreasing optical luminosity (Section 4.1 (ii)). {\em In Paper II we had shown that in Seyfert 
1s, IR loudness anti-correlates with optical luminosities and galaxy sizes and we had suggested it
is a measure of dust extinction and host luminosity. Here we show that this is the effect of 
interacting members in this class, while  isolated Seyfert 1s have a large range of optical and IR
luminosities.}

\section{Characteristics of the Encounters}

In the previous section we have shown that some of the important differences between Seyfert 
type 1 and 2 galaxies are strongly related to their (different) environmental properties, 
providing clear evidence against the simple orientation model often postulated for 
Seyfert galaxies. It is thus of great interest to determine what 
environmental factors influence the (observed) nuclear Seyfert type.   
Moreover, we have seen that Cold galaxies have markedly different environmental properties 
than those of the Warm sample. However, it is not clear whether an IR-Cold galaxy {\em will} develop 
nuclear activity
or, alternatively, what is the process responsible for ``warming'' the mid-IR colours of interacting
Seyfert galaxies.
In this section we shall address these questions by attempting to quantify the {\em close}
environmental properties of the non-isolated objects in all our samples.
These include 8 Seyfert 1s, 20 Seyfert 2s and 12 Cold galaxies.

In Section 2 we have defined the relevant quantities, $Q_{D}$, $M_{s}-M_{p}$, $D_{s}/D_{p}$,
$S/D_{p}$. In Figure~\ref{f3} we plot their distributions
for the brightest and the closest companions, respectively.
The median values and standard deviations of these distributions are given in 
Table~\ref{tab2}. Before discussing the results we draw attention to the
Seyfert 1 galaxy, IRAS 02253-1642, which appears to be a strongly interacting system with tidal
morphology. However, closer examination shows that one of the two objects is compact and star-like
while the other is highly elongated and curved, suggesting that this might be a lensed background
galaxy. This object is included in the Seyfert 1 samples of a number of recent studies but no 
individual information is given for it. In what follows, we include it in our histograms
and plots, but we also discuss the effects of its inclusion in our statistical results (when 
relevant).

\placetable{tab2}

\placefigure{f3}

\subsection{Brightest Companions}

\subsection*{\it Separation}

Let us first consider the brightest companions in all samples, selected as described in 
Section 2 (four upper panels of Figure~\ref{f3}). The projected separations $S$ normalized by the diameter of the primary $D_{p}$
span a large range for the Warm sample, with a median value of 1.2 (median projected separation 
$\sim$40 kpc) for the Seyfert 2s and 1.57 (median projected separation $\sim$29 kpc) for the 
Seyfert 1s. In fact, the distribution of Seyfert 1 bright companions is rather flat, showing a 
similar probability for a galaxy to have a companion at 1$D_{p}$ or 4 $D_{p}$, while the Seyfert
2 distribution is more segregated between 1-2$D_{p}$. The Cold galaxies show a very
different distribution of separations, peaking at much smaller values
with a median 0.56 (observed separation $\sim$22 kpc). The F and K-S tests show significantly 
different Warm and Cold galaxy distributions. These results do not change significantly if IRAS
02253-1642 is excluded from the Warm Seyfert 1 sample.

The smaller separations found for the Cold sample agree with our previous results that it
contains almost exclusively strongly interacting systems. The difference in mean separations found
between Seyfert 1s and 2s is expected, given the larger fraction of disruptive encounters found 
among the latter. The median relative separation in Dahari's Seyfert sample was 
$S/D_{p}$=1.4 (\cite{dahari84}), comparable to the median for the whole Warm sample. For all our
objects (Warm and Cold) the absolute projected separations were $\leq$120 
kpc, which is approximately the typical search radius in most of the recent studies of (optically 
selected) Seyfert environments. 

\subsection*{\it Companion ``Mass''}

\begin{figure}
\epsscale{1.}
\plotfiddle{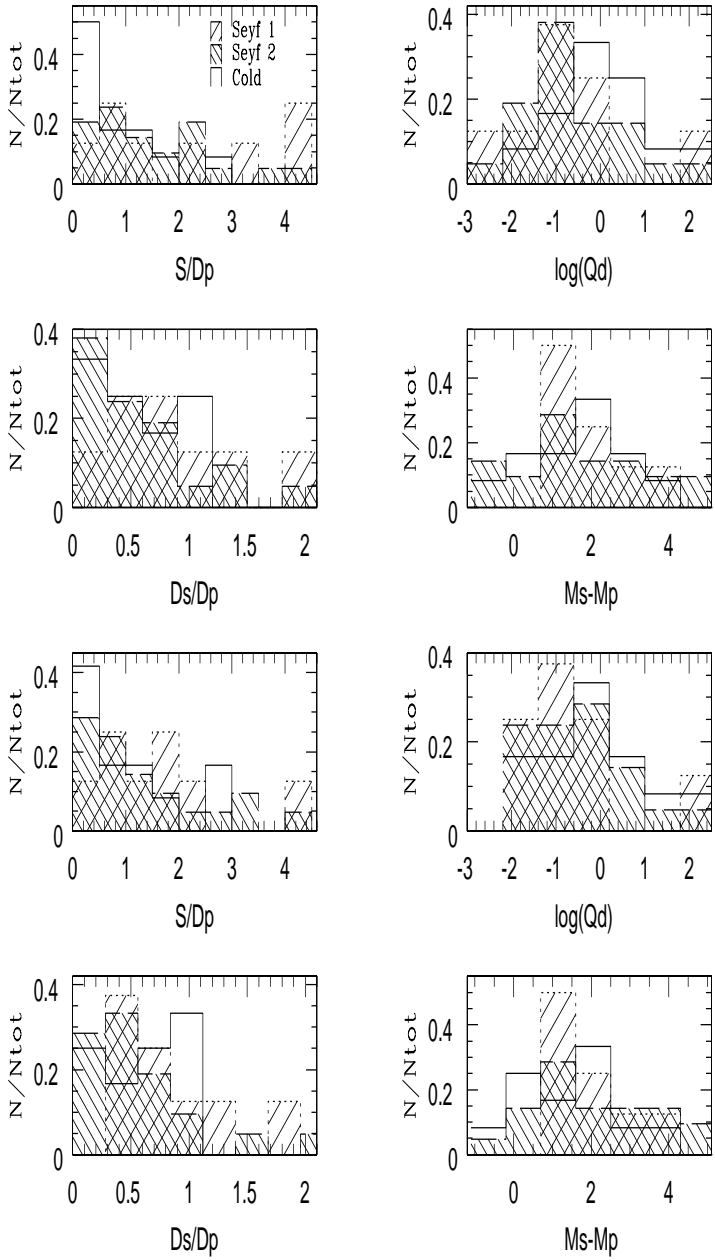}{400pt}{0}{50}{90}{-160}{-150}
\caption{Distributions of projected separations normalized to the diameter of the primary ($S/D_{p}$), interaction strength ($Q_{D}$), diameter ratio (D$_{secondary}$/D$_{primary}$) and magnitude difference (M$_{secondary}$-M$_{primary}$) for our three (sub)samples (see text for definitions). The four upper panels are plots for the brightest companions and the four lower panels for the closest companions, within 3$\times D_{B25}$. \label{f3}}
\end{figure}

Galaxy separation is not the only factor that determines the strength of an interaction, the 
tidal force being also proportional to the relative mass of the companion (Section 2). Since we 
have no direct ``mass'' measure, we can use instead the relative sizes and/or luminosities of the 
two galaxies. In Figure~\ref{f3} we show the distribution of major axis ratios and magnitude
differences between companion and primary galaxy. We find that the relative sizes of the
Seyfert 1 and 2 companions are quite similar, while the Cold companions have the tendency to be 
larger. However, the three distributions are statistically similar, according to the F
(variances), Student's t (means) and K-S tests, with medians in the range 0.5-0.8. 
\cite{fuentes88} imposed a lower limit to the companions diameters 
$D_{s}\geq$0.25$D_{p}$, in order to be able to disentangle stars from galaxies on the POSS 
plates. From Figure~\ref{f3} we see that a similar limit would have caused us to miss 
$\sim$30\% of the Seyfert 2 and Cold galaxy companions. The companion relative brightnesses, 
shown in Figure~\ref{f3}, are also similar for the two Seyfert subsamples, while the Cold 
companions are shifted to somewhat fainter magnitudes (that is, larger $\Delta$m). However, the 
three distributions do not differ at a statistically significant level and their medians are in 
the range 1.55-1.8 mag. \cite{rafanelli95} have imposed a selection criterion more stringent 
than ours, $\Delta$m(max)=3 mag, based on the distribution of relative companion magnitudes in 
the Vorontsov-Velyaminov {\em Atlas and catalogue of Interacting Galaxies} (\cite{vorontsov59}). 
From Figure~\ref{f3} we see that such a limit would cause us to miss $\sim$20\% of the companions
in each sample.

\subsection*{\it $Q_{D}$}

Let us now consider the interaction strength, $Q_{D}$, that includes both companion mass and 
separation. We find that Seyfert 1s tend to have a tail to lower $Q_{D}$ values (in particular if 
we exclude the uncertain galaxy mentioned earlier) and Seyfert 2s a tail to higher values. 
The median value of $Q_{D}$ is 0.12 for both subsamples (0.10 for the Seyfert 1 subsample, if we exclude
the one ambiguous object discussed earlier). The Cold galaxies have significantly larger $Q_{D}$ values, with a 
median $Q_{D}$=1.44. The K-S test shows that the null hypothesis, that two samples are drawn from
the same parent population of $Q_{D}$ values, is confirmed for the Seyfert 1 and 2 subsamples and 
rejected for the Warm and Cold samples, at significance levels $\sim$95\%.

Dahari has found that the fraction of Seyfert/Normal galaxies in his sample is 
$\sim$7 when $Q_{D}\geq$1 and drops to 2 when 1$<Q_{D}\leq$0.05, which suggests that Seyferts
are more likely to have close and/or more massive companions to perturb them. The median $Q_{D}$
value in his Seyfert sample was 1.18; he suggested that $Q_{D}\geq$1 defines ``strong'' 
interactions. 
\cite{byrd86} showed that tidal triggering can be a physically
{\em sufficient} mechanism to induce nuclear activity, through the formation of bar or spiral 
instabilities and consequent mass inflow into the nuclear regions. They modeled the gravitational
instability flows in terms of $Q_{D}$ and $Q_{Ostriker}=Q_{O}=$ Halo/Disk mass and found such
instabilities to set in for $Q_{D}\geq$1 independently of $Q_{O}$, while for $Q_{D}\geq$0.05
only for low halo systems. Among our Warm Seyfert galaxies with detected companions, we find a 
similar fraction of types 1 and 2, 25-30\%, to have $Q_{D}\ge$1 this fraction increasing to 
$\geq$50\% for the Cold sample. 

We conclude that, {\em if the presence of bright companions 
within a radius of three galaxy diameters dominates the environmental effects in IR-luminous 
galaxies, our results indicate that the interaction strength (and in particular the companion 
proximity) is an important discriminator between Warm and Cold samples}. 

\subsection{Closest Companions}

Let us now consider the properties of the closest companions for the three (sub)samples (four lower
panels of Figure~\ref{f3}). We find shifts
but no major differences in the companion relative ``masses'' (that is, sizes and magnitudes) between
Warm Seyfert types 1 and 2. 
However, we find significant difference in the distribution of projected separations between these two
subsamples: for Seyfert 2s, the separations are shifted and peak at smaller values. 
On the other hand, the Seyfert 2 and Cold galaxy companions have similar relative projected 
separations (the K-S test shows that the null hypothesis of matching distributions is
confirmed at the 97\% significance level).
The most important effect is depicted in the interaction strength, $Q_{D}$, distributions:
the two Seyfert subsamples have very different distributions and means (in particular when the 
ambiguous Seyfert 1 galaxy, discussed above, is excluded). On the other hand, the $Q_{D}$ 
distributions  are very similar for the Seyfert 2 and Cold (sub)samples (although the K-S statistical
significance remains less than 95\%). The different $Q_{D}$ distributions are likely to reflect the 
denser environments of Seyfert 2s (and Cold galaxies) compared to S1s. 
We thus conclude that {\em in Warm non-isolated Seyferts, the proximity of close companions seems 
to be the best discriminator between their nuclear activity types}.

\subsection{Correlations between Interaction Strength and IR Properties}

\begin{figure}
\epsscale{1.}
\plotfiddle{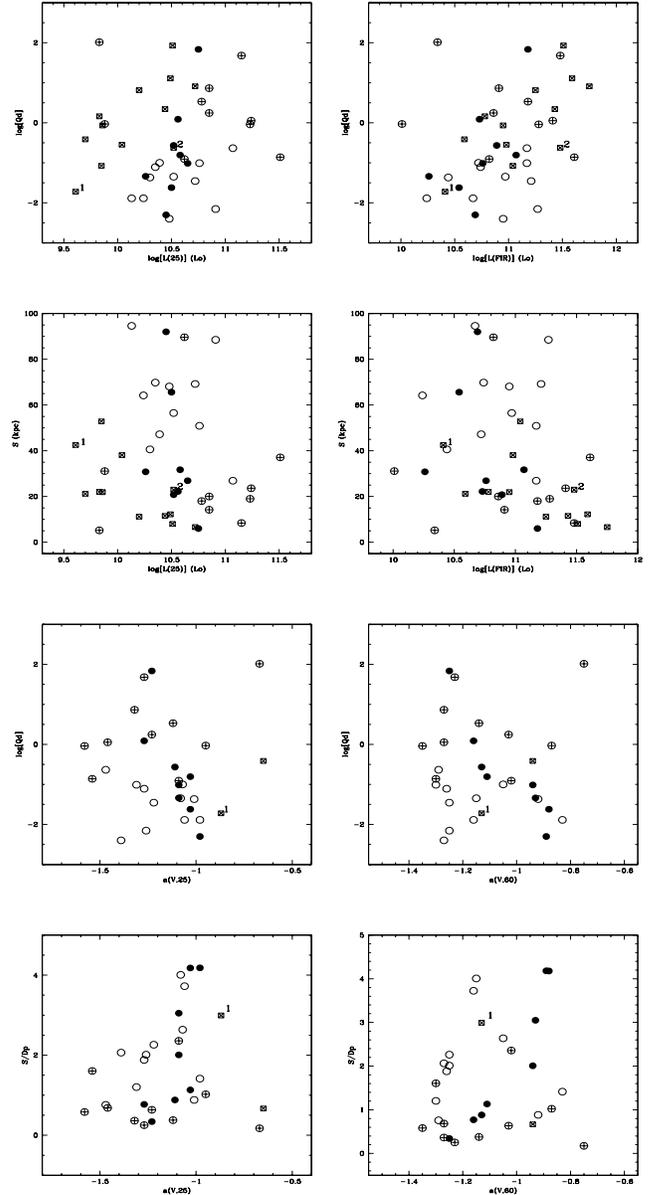}{400pt}{0}{60}{60}{-180}{-10}
\caption{IR properties (log($L_{25}$) and log($L_{FIR}$) (in solar units \Lsun) and IR-loudness indices $\alpha_{(V,25)}$, $\alpha_{(V,60)}$versus projected separation $S$ (in kpc) or $S/D_{p}$ (scaled to the size of the target galaxy) and interaction strength log($Q_{D}$), for our three (sub)samples. Symbols are as in Figure 2 (crossed open circles indicate strongly interacting Seyfert 2s). The labels {\em 1} and {\em 2} indicate the nuclear types of the Cold sample Seyferts. \label{f4}}
\end{figure}

Now we shall investigate how the interaction strength correlates to the IR
properties in our samples. The main correlations found are shown in Figure~\ref{f4}.
(The Seyfert 1 galaxy with the largest $Q_{D}$ and smallest $S$ in this figure, is the uncertain 
object mentioned earlier and it is detached from the rest of the sample in most diagrams.)
In order to better understand our correlations, we denote with crossed circles the strongly
interacting Seyfert 2 galaxies (interaction class $S$) and with open circles those in interaction
class $C$. Our main conclusions are:

\placefigure{f4}

(i) {\em Upper four panels}: For Cold galaxies the IR luminosities, both $L_{25}$ and $L_{FIR}$, 
correlate with 
interaction strength (the Spearman's rank-correlation test gives correlation coefficients 0.60 
and 0.67 at 0.02 significance level) and projected separation (correlation coefficient -0.61 at 
0.03 significance level), in the sense that galaxies with closer companions and larger $Q_{D}$ 
tend to be IR-brighter. This correlation holds better when the brightest companions are considered. 
There is a similar tendency for the strongly interacting Seyfert 2s to have 
larger IR luminosities, compared to the rest of the Seyfert 2 subsample and to the Seyfert 1s. 
Moreover, Figure~\ref{f4} shows that for similar interaction strengths, 
Warm Seyferts of both types have larger $L_{25}$ than Cold galaxies, while Warm and Cold 
samples overlap in the $L_{IR}$ vs $Q_{D}$ diagrams. These results thus support the conclusions
that we reached in 
Section 4, from different parameters: {\em the excess $L_{25}$ in Warm Seyferts is mainly due to the
AGN contribution, although in strongly interacting Seyfert 2s it is further enhanced by star 
formation. In Cold galaxies, interaction induced star formation is the source of IR activity at all 
wavelengths}.

(ii) {\em Lower four panels}: In Seyfert 1s the IR-loudness coefficients correlate with interaction 
strength and anti-correlate with relative separation (in particular when their brightest companions 
are considered). The Spearman's rank-correlation test gives a correlation coefficient of -0.9 to
-1.0 with significance better than 0.005 for IR-loudness vs $log(Q_{D})$  and a correlation 
coefficient 0.9 with significance better than 0.004 for IR-loudness vs separation. {\em We have 
seen previously that the IR-loudness in Seyfert 1s correlates with interaction stage and here we 
find that within similar interaction stages it scales further with interaction strength}.

(iii) We searched for correlations between the IR colour indices $\alpha_{(25,60)}$, 
$\alpha_{(60,100)}$ and $Q_{D}$. We find no correlation for any of the samples. This is a 
significant result because 
it indicates that {\em although the strength of tidal perturbations is a discriminator between 
Warm and Cold samples, the mid-IR excess (or lack of) is not a simple function of interaction 
strength} (a conclusion reached in Section 4.2 from different parameters). 

\subsection{Cold Seyferts}

At this point it is interesting to consider the properties of the four Seyfert galaxies
in our Cold sample. Two of them, type 2 IRAS 04265-4801 (interacting) and type 1 IRAS
06506+5025 (barred galaxy with companion; labeled {\em 1} in Figure~\ref{f4}) show the lowest 
IR luminosities among the Cold sample. In Paper II we have seen that they are also faint in 
optical wavelengths (among the faintest compared to the respective Warm Seyfert subsamples) and have
very red colours, at all spatial scales. Moreover, they both have the smallest $Q_{D}$ compared 
to the rest of the Cold sample. Although they harbour a Seyfert nucleus, it is likely
that large amounts of cooler dust in their disks or/and an intrinsically faint AGN 
are responsible for their cold mid-IR colours. The first option is certainly the case for 
IRAS 04265-4801 which suffers large extinction so that even its nucleus cannot be 
unambiguously identified (see also Paper IV). IRAS 06506+5025 has a rather complex colour 
distribution (Paper IV) and both the above factors are likely to be responsible for its 
inclusion in the Cold sample. The other two Cold Seyferts, IRAS 23128-5919 (merger) and IRAS
19184-7404 (strongly interacting), both of nuclear type 2, are among the most IR-luminous objects in the 
Cold sample. The latter object is shown in Figure~\ref{f4} (labeled {\em 2}) to have similar 
interaction strength and $L_{25}$ as the Warm sample. The cold 
25-60 $\mu$m colours of these two galaxies result from the fact that their far-IR luminosities 
are also enhanced (compared to the rest of the Cold {\em and} the Warm sample). Thus, 
unusually large amounts of dust and/or strong star formation in their disks could be invoked 
to explain their mid-IR colours, although we have no direct evidence for this, as we lack 
photometric data for both objects. Although such a small number (four) of cases is not enough to
establish any firm conclusions, the {\em existence} of these very different
cases of Cold Seyfert galaxies indicates that {\em the occurrence of a mid-IR excess must be 
related to some intrinsic properties of the host galaxy rather than being merely a transition 
period in the evolution of strongly interacting systems}. The evolutionary interpretation could 
account for the fact that Cold galaxies show statistically larger $Q_{D}$ (Section 5.1) and might 
indeed be the case for the two most powerful Cold Seyfert 2s. However, the existence of the other 
two Cold Seyferts shows that effects related to the properties of the progenitor 
galaxy (such as gas or dust content) and/or the geometry of the interaction, must be also
important for the development or lack of a mid-IR excess.

\subsection{Mergers}

The merger systems in our samples (interaction class (Mg)) deserve special mention. These are
systems possessing double nuclei within a common envelope, or merger remnants with two tidal
tails emanating from the same body in opposite directions.
There are three double-nucleus mergers in our Warm sample, one Seyfert 1 and two Seyfert 2s,
and one double-nucleus merger in the Cold sample which is also spectroscopically classified as a
Seyfert 2. Furthermore, the Warm Seyfert 2 and Cold subsamples contain one advanced merger each: 
in the first case we can hardly identify the
two nuclei in continuum light but they are better visible in $H\alpha$ emission; in the second,
the two nuclei are not any more distinguishable but the remnant has the characteristic two-tail 
appearance of a recent merger. In Table~\ref{tab3} we list some of their properties and
their $R$-band images are shown in Figure~\ref{f5}. 

\placetable{tab3}

\placefigure{f5}

We find that the three Warm Seyfert 2 mergers, independently of coalescence state, have larger IR 
and optical total and disk luminosities and warmer 60-100 $\mu$m colours compared to the Seyfert 1
merger case, but also fainter nuclei and redder optical colours. These properties indicate that
there is a larger fraction of warm dust in the {\em nuclear} regions of the Seyfert 2 
mergers. The Cold Seyfert 2 double nucleus merger shows similarly large IR luminosities and warm
60-100 $\mu$m colours (no optical photometry available) as the Warm Seyfert 2 cases.
\cite{murphy96} found that 47\% of their ULFIR sample show double nuclei and
suggested that this must be typical for the ultraluminous IR phase (whose lifetime is
estimated to be 2$\times$10$^{8}$-10$^{9}$ yr). We indeed see from Table~\ref{tab3} that all our 
merger systems, with the exception of the Warm Seyfert 1, have $L_{FIR}\geq$10$^{11}\Lsun$.
In Paper IV we have shown that the four {\em double nucleus} mergers have remarkably similar 
colour and line emission morphologies: one of the two nuclei is brighter and redder, the line 
emission being centered on it with an extended morphology in direction perpendicular to the 
line connecting the two nuclei. In the case of IRAS 13536+1836 it was shown that the brightest 
nucleus is also the activated one (\eg \cite{eleni1}); the same object is optically classified 
as a Seyfert 2 but shows Seyfert 1 characteristics in polarized light. Given their common 
properties, it is likely that the other three double nucleus mergers are similar objects
to IRAS 13536+1836.

The evidence for larger extinction in the Seyfert 2 mergers suggests that their optical 
spectral classification is most likely to be affected by obscuration. If there is an evolutionary
sequence between the Seyfert 1 and 2 types, as we suggest later in this paper, {\em the Seyfert 2 
merger cases might be representing this rare phase of transition from nuclear type 2 to type 1}.
The small number of (six) objects examined here and their varied properties consist circumstantial
evidence but do not allow to further establish any firm conclusions.
A follow-up project to map the detailed morphologies, kinematics and ionization of selected double
nucleus mergers is under way.

\section{Discussion}

\subsection{Environments and Interactions}

Whether there is a preferential occurrence of interacting/merging systems and morphological 
distortions in Seyfert galaxies is a controversial issue and, despite the large number of related 
studies, non conclusive. There have been several recent reviews of the statistical properties of 
Seyfert galaxy environments (\eg \cite{laurikainen95,hacyan99}). Such studies 
involve mostly optically selected samples and are mainly of two types: they involve studying either 
the frequency of AGN
occurrence in morphologically disturbed galaxies or the frequency of disturbed host morphologies 
and/or nearby companion galaxies in nearby AGNs (\eg Seyferts). In the first approach, no excess 
of Seyfert nuclei (and often even a lack of them) is found in disruptive systems; a marginal excess of Seyferts is found among systems
at lower interaction levels (\eg \cite{keel85,dahari85,bushouse86,sekiguchi92}). In the second 
approach the results are more contradictory: an
excess of companions for Seyferts compared to normal galaxies has been reported 
(\eg \cite{dahari84,mackenty89,monaco94,rafanelli95}) but other studies
indicate that Seyfert and control sample environments are comparable 
(\eg \cite{fuentes88,laurikainen95,robertis98a}). In fact, the situation 
seems even more complicated as sometimes excess of only faint (\cite{fuentes88}) or on the 
contrary only bright (\cite{hacyan99}) companions is claimed. The reasons for these 
discrepancies are usually attributed to differing definitions of ``environments'', data 
limitation in detecting companions, biased working samples and 
lack of or poorly defined control samples. The numbers for the apparent companion excess in
Seyfert samples vary widely. \cite{dahari84} found that 40\% of his Seyferts have companions, 
compared to only 23\% of the control sample and he suggested that the excess of companions among 
Seyferts corresponds to large interaction strengths $Q_{D}\ge$1. \cite{byrd87} showed that the 
fraction of Seyferts with companions in the Dahari sample could be as large as 50\%, when account is 
taken of companions not 
considered, due to the limiting magnitudes and maximum search radius imposed by Dahari. 
\cite{rafanelli95}, on the other hand, found that each of the two Seyfert 
types show excess of companions up to $\sim$12\%, compared to a maximum of 5\% for the control samples.

When Seyfert 1s and Seyfert 2s are considered separately a much clearer environmental connection
appears. In general, the relative densities in Seyfert 2 environments are 
found to be 1.6-2.7 larger than those of normal galaxies, while Seyfert 1s have identical 
or smaller companion densities compared to normal galaxies (\cite{laurikainen95}). 
The larger density of Seyfert 2 companions seems moreover to depend strongly on morphological 
type (\cite{petrosian82,mackenty89,hacyan99} and references therein).  
Different host morphological types are often postulated, with type 1s residing in earlier type 
hosts than type 2s (\eg \cite{malkan98}). However, these results are contradicted by 
other recent studies. \cite{mackenty90} reports that spiral structure does not seem 
predominant amongst his Seyfert sample (as postulated by \eg \cite{simkin80,dahari84}), although 
he finds that in general Seyferts have axial ratios $\leq$0.5 (see also \cite{keel80}).
\cite{maiolino95} found no significant differences in the Hubble types of Seyfert 1 and 2 
galaxies and concluded that the enhanced star formation activity, often postulated for Seyfert 2s,
does not seem to be related to the prevalence of late-type disks among them.

How do our results compare to those of previous studies?
The fraction of systems with companions is generally larger among our IR-Warm Seyfert sample
compared to the optical Seyfert samples: 43\% for the Warm Seyfert 1s and 63\% for the Warm Seyfert 
2s. As we mentioned earlier, this would be consistent with the supposition that IR-activity does probe
strong interactions. More importantly, 
we find twice as many Seyfert 2s than Seyfert 1s in strongly interacting/merging systems, 
in rough agreement with the ``optically-based'' results. We also find (Paper III) a tendency for Warm 
Seyfert 1s to reside in earlier type hosts (peaking around SO-SO/a) compared to Warm Seyfert 2s 
(peaking at Sa types). Among our Warm Seyferts with companions we find a smaller fraction than
in Dahari's sample,  25-30\%, to have $Q_{D}\ge$1 (a {\em similar} fraction for types 1 and 2)
this fraction increasing to 50\% for the Cold (mostly non-Seyfert) sample. Following our results
described in Sections 3-5 we put forward the following suggestion: {\em although the evidence for a 
causal connection between galactic interactions and nuclear activity in our sample is substantial,
the development of nuclear activity and its observability depend on the time evolution since the 
last interaction, as well as on the host intrinsic properties}.

We now address the question of whether important differences exist between isolated Seyfert 1 and 
2 galaxies.
In optical samples, the fraction of solitary Seyferts ranges between 10-25\%. This is 
comparable to the fraction we find for Warm Seyferts (similar for type 1s and 2s) if we
consider undisturbed, non-barred/ringed systems. However including the latter, the fractions
of isolated systems increase by 1.5 and 3 times for Seyfert 2s and 1s respectively, that is, 
{\em the majority of Warm Seyfert 1s lie in ``isolated'' hosts with disturbed morphologies}.
This suggests that these Seyfert 1s are only ``pseudo-isolated''.
Multiple tidal perturbations from faint companions (too faint to be considered as such) could 
be responsible for generating non-axisymmetric potentials such as bars or rings and in fact
this process was previously invoked to explain the significant fraction of isolated Seyferts 
observed by \cite{fuentes88}. Alternatively, the apparent isolation of Seyfert 1s in our sample
could be a 
long-lasting effect of encounters, in the sense that dynamical friction in their vicinity has 
gradually led to removal of companions (\eg \cite{laurikainen95}). 
Whichever of these two scenarios - different environments or evolutionary effect - applies,
our data show that {\em there is a fundamental difference between the Seyfert 1 and 2 ``isolated''
galaxies in our sample}. We favour the evolutionary effect
for one more reason: an important conclusion in this paper is that, among non-isolated 
Seyferts there is no significant difference in the relative size and luminosity of companions
between the two types (Sections 5.1 and 5.2). In other words there is no reason to postulate 
different, more disruptive environments for Seyfert 2s.

\subsection{Morphologies}

The question arises: how common are disk instabilities and what do they imply for the 
recent interaction history of a galaxy? From a theoretical point of view, the modeling 
of interactions/mergers between two disk/halo systems of comparable mass shows commonly tidal 
triggering of central bars, that perturb the orbits of gas and stars in the host galaxy causing 
infall and circumnuclear star formation, with probable subsequent AGN feeding/triggering (\eg
\cite{byrd86,barnes96,mihos96,mihos99}). Although earlier observational 
work indicates a prevalence of kpc-scale distortions (bar/ring formations) in Seyferts 
(\cite{simkin80}), according to the most recent optical and IR studies the frequency of occurrence
of such features among Seyferts (25-30\%) is similar to that among normal spirals (\eg
\cite{mackenty90,mulchay97,robertis98b}). In fact, a few workers find that
the presence of bars {\em inhibits} AGN segregation (\eg \cite{monaco94}). This observational 
evidence suggests that either bars are not a universal fueling mechanism, that the bar
formation inhibits gas infall (stabilizing the disk) and thus AGN triggering, or that bars tend 
to be destroyed after the formation of a black hole (\eg \cite{hasan90,pfenniger90}).
We believe that the lack of a significant excess of bars in AGNs is connected to 
an evolutionary scenario for Seyferts. In our sample of Warm Seyferts, if we exclude strongly 
interacting systems, we find a similar fraction of barred/ringed galaxies among Seyfert 1s and 2s.
However, the former tend to be purely barred/ringed and to reside in isolated hosts, whereas the 
latter can have some other tidal feature as well and most of them have at least one detectable 
companion. Moreover, all the barred/ringed strongly interacting Warm Seyferts are type 2. We believe
that these numbers reflect differing interaction {\em stages} for Seyfert 1s and 2s, 
while at the same time indicate an equal probability for Seyfert 1s and 2s to  
develop bar or ring instabilities. It is of particular interest that bar/ring and tidal 
features seem to be mutually exclusive in Seyfert 1s (only one possible candidate), while their simultaneous
appearance is common in Seyfert 2s (especially when strongly interacting systems are 
considered). This most likely reflects differing evolutionary lifetimes for the bar 
and tidal features and differing rates of bar vs tidal formation in different 
types of interactions (see also Section 6.4).

\subsection{IR Activity and Star Formation}

If interactions and Seyfert activity are linked by an evolutionary sequence, we can
ask whether starburst and Seyfert activities are similarly linked.
 The ratio of Seyfert to starburst galaxies was found to be
a function of IR power and was suggested to be also a function of time since a merger occurred 
(\eg \cite{veilleux99,sanders99} and references therein). It has been proposed that AGNs and
starburst galaxies are linked according to the following scenario:
Interactions or mergers of gas-rich spirals funnel gas into the nuclear regions, triggering 
intense star formation that produces the bulk of IR luminosity in the initial stages of the
merging process. When the gas becomes more concentrated to the center ($\leq$1 kpc)
it feeds the central black hole and triggers an AGN that dominates the luminosity output of the 
system
and this corresponds to the peak IR-phase. As time goes by, superwinds from the newly formed stars
will clear away interstellar material from the central regions, so that most of the circumnuclear starburst eventually stops,
while the AGN fueling continues for a while. During this phase, the AGN becomes
visible and warmer IR colours might develop, this primarily depending on the small scale (torus) 
geometry (for detailed description of similar scenarios and relevant references see the review 
papers of \cite{sanders96,sanders99}). 

A similar evolutionary scenario was suggested by \cite{hutchings91}, studying the interaction 
properties of a sample of IRAS galaxies with a range of IR and nuclear activity types. 
They found that the steep spectrum ($\alpha_{(25,60)}\leq$-1.3) galaxies (whether these are HII,
LINERs or Seyfert 2) are the younger and stronger interactors and show larger dust obscuration.
On the other hand, the flat spectrum Seyfert 1 and Seyfert 2 galaxies appear to be intermediate 
age/strength interactors. The authors suggest that the IR-warmness possibly represents different
evolutionary paths of active galaxies. The steep-spectrum Seyfert 2s might be at an earlier 
evolutionary stage than the flat spectrum Seyferts, the type 2s being hidden type 1s at the 
``flat-spectrum stage''. Whether the steep-spectrum Seyfert 2s will evolve to flat-spectrum Seyfert 2s
or to steep-spectrum low-luminosity AGNs (such as LINERs) is unclear.
Using polarimetric studies to detect hidden broad line regions (BLRs) in Seyfert 2 galaxies, Tran (1993, 1995a, 1995b, 1995c)
 and \cite{heisler97} are led to somewhat different conclusions:
hidden (BLRs) are found predominantly in interacting galaxies with warm mid-IR colours 
($\alpha_{(25,60)}\leq$-1.6) and smaller internal extinction. Within the orientation model, this
would mean intermediate (between pole-on and edge-on) torus viewing angles for the flat-spectrum 
(warm) Seyfert 2 galaxies. It would also mean that interactions influence the visibility of the BLR,
by either producing the scattering ``mirrors'' that allow detection of the hidden Seyfert 1 or on
the contrary by producing the circumnuclear obscuring material that hides the BLR of the Seyfert 1
nucleus. In fact, \cite{tran95c} suggests that the obscured Seyfert 2 nuclei might represent an
evolutionary stage between ``pure'' (unobscured) Seyfert 2 and Seyfert 1 type galaxies.
This was inspired by the evolutionary scenarios connecting starburst and AGN activity, suggested by 
\cite{osterbrock93}. According to these, interacting galaxies pass through an initial starburst phase
before the AGN dominates their power output: at earlier stages the galaxy appears as an ultraluminous Seyfert 2 galaxy,
later evolving to a Seyfert 1 (unless viewed edge-on), then to a lower luminosity AGN and finally
to a normal spiral as the nuclear activity ceases (possible variations to this scenario are also
suggested). Finally, \cite{sanders99} notes that most highly IR luminous ($L_{FIR}\sim$10$^{13}\Lsun$)
galaxies, were found to be dusty Seyfert 2s that later on were shown to be obscured Seyfert 1s (or, 
at those luminosities, QSOs). 

Our study of the IR Warm and Cold samples, selected according to the shape of their spectra between 
25 and 60 $\mu$m, probes such possible evolutionary link between starburst-dominated and AGN-dominated
IR emission.  
In Papers II-V we showed that in Cold galaxies the bulk of their IR emission is dominated by warm dust
in their disks, powered by strong star formation. In their large majority, these systems undergo 
disruptive
encounters. Their IR emission at all wavelengths increases with interaction strength and the 25-60
$\mu$m colours become colder at smaller projected separations. Furthermore, we found a transition 
in most observed properties, from the Cold sample to the Warm Seyfert 2 and to the Seyfert 1 samples: 
brighter nuclear magnitudes, bluer nuclear optical colours,
bluer 12-100 $\mu$m colours, decreasing IR-excess and IR loudness, increasing companion
separations and decreasing interaction strengths. This transition could be further expanded for 
their structural (bulge, disk) parameters and host morphologies (Paper III). 
Our observational results thus, generally sustain an evolutionary scenario, such as 
the ones described above, between the Cold and Warm samples that offers a possible explanation for 
their differing dust temperatures and dominant activity type.
However, the occurrence of four Seyfert galaxies in our Cold sample, showing a variety of 
optical and IR properties, indicates that other factors, such as the gas and dust content of the 
progenitors and maybe also the interaction geometry, are important in determining whether
the galaxy will go through an IR-Warm phase. In what follows we will argue for a similar 
evolutionary connection also between the Warm Seyfert 1 and 2 types, as evidenced from their
strongly differing properties, described in Papers II-V. 

Although star formation seems to be the dominant source of far-IR emission (longwords of 60
$\mu$m) not only in starburst but also in Seyfert galaxies, observational evidence has 
accumulated suggesting that the mid-IR emission originates near the nucleus and is a 
combination of nuclear and starburst thermal components (\eg \cite{bonato97,rodriguez97,sanders99}).
The relative contributions of these two components are postulated to be related
to the type of Seyfert activity, with highest dust temperatures for the type 1s. 
A number of important results relevant to these questions were demonstrated in Papers II-V: 
(i) Warm Seyfert 2 disks tend to be brighter than Warm Seyfert 1 disks at optical wavelengths and 
their mid- and far-IR properties correlate with total optical luminosities and host galaxy size, 
indicating that the bulk of Seyfert 2 IR emission originates in their disks.
(ii) Optical colours indicate that the Warm Seyfert 2s are {\em overall} more dusty than the Warm
Seyfert 1s, their nuclei suffering the highest obscuration. (iii) The 25 $\mu$m emission in Warm
Seyferts is primarily due to warm dust heated by the AGN, but 
in strongly interacting systems (and thus, primarily in Seyfert 2s) it is further enhanced by disk
star formation. (iv) The IR luminosity longwords of 60 $\mu$m indicates mainly
excess emission at large scales and is similarly enhanced by strong interactions in the Warm and 
Cold samples. 

The importance of disk and nuclear starbursts specifically in Seyfert 2s is supported by a 
variety of observational studies covering a large wavelength range: 
strong far-IR and CO emission by cool dust (\cite{heckman89}), strong extended mid-IR emission 
and spectral features (\cite{maiolino95}), large n-IR L/M ratios (\cite{oliva95}), optical and
UV spectral features and imaging 
(\cite{fernandes95,heckman95,heckman97,rosa98,duilia99,rosa99,heckman99}).
On the basis of near-IR colours and colour gradients, \cite{hunt97} suggested the contribution of 
an intermediate-age stellar population ($<$10$^{9}$ years) in the Seyfert 2 disks, while mainly
old stars are found in Seyfert 1s. \cite{fernandes98} and \cite{schmitt99} argued that most of the optical and
near-UV featureless continuum in Seyfert 2s is produced by stars younger than 10$^{8}$ years.
\cite{fernandes95} suggested that the circumnuclear star formation in these objects is associated
with the obscuring molecular torus. \cite{maiolino97} have argued that the larger {\em disk} 
star formation rate found in Seyfert 2 galaxies is not likely to be due to larger molecular gas 
content but to rather more efficient star formation from similar gas amounts, which they attribute to 
the larger occurrence of distortions found in Seyfert 2 disks.
In Paper IV we have shown ample evidence for the occurrence of strong star formation in the 
Seyfert 2 disks: (i) colour and emission line distributions that suggest dust extinction
associated with on-going star formation, mostly in spiral and tidal features (ii) negative radial
colour gradients that correlate with interaction strength and IR luminosities (see present 
paper) (iii) large star formation rates as deduced from their IR emission longwords of 60 $\mu$m
(iv) starbursts of 1-0.5 Gyr or younger, superposed on the older underlying galaxy population,
in particular for strongly interacting systems. Most of these properties are shared with the
Cold galaxies in our sample, while Warm Seyfert 1s show often opposite colour gradients and 
mostly older stellar populations, although in a few cases there is evidence for circumnuclear 
star formation.
The plausibility of the evolutionary scenario in which strong interactions are responsible for 
both enhanced star formation and activation of the galactic nucleus, suggests that 
Seyfert 1 and 2 galaxies might represent different phases in an evolutionary cycle.

\subsection{An Interaction Sequence}

The observational evidence put forward in the previous section does not sustain the simple
orientation/obscuration picture for Seyfert type 1s and 2s, although it might be true for {\em some} 
of the Seyfert 2s. Here we suggest an alternative view that takes into account {\em the time evolution
of environmental and morphological properties of Seyferts}. The failure to consider these factors
in most Seyfert environmental studies are likely to account for the discrepancies in their 
results: most of the criteria used in the search for companions favour close environments 
(typically few times the diameter of the target galaxy), relatively bright companions and a 
particular interaction phase when the two systems are bound but not yet merged. Moreover,
environmental studies often are not coupled to morphological examination and thus might be
missing (i) faint tidal features that indicate an evolved merger, or tidal 
perturbations due to a faint companion (or a companion lying outside the maximum search radius) 
(ii) bars and rings that suggest perturbation by smaller companions or at earlier epochs
(iii) central elongations and even faint secondary nuclei that indicate a recent merger.

Numerical simulations have indeed shown that the minimum interaction strength needed to trigger
disk instabilities can be reached (i) during the perigalactic passage of very small companions, 
(ii) by larger companions at greater distances ($\sim$30$D_{p}$ or more), or (iii)  in massive 
hyperbolic encounters where shortly after perigalacticon, $\sim$10$^{8}$yr(\eg \cite{byrd86}),
the companion reaches large distances. At perigalacticon, tidal distortions are triggered
that are quickly amplified by disk self-gravity into bar or spiral instabilities.
In these cases rapid gas inflow leads to an early episode of 
starburst or AGN activity well before the galaxies merge (within a few rotation periods). 
At this stage the galaxies have moved further apart and one can observe disk tidal 
distortions and perhaps an inner bar.
This could be the case in several Seyfert 2 galaxies in our sample, where strong tidal features 
and central bars are simultaneously observed. 

It is estimated that normal galaxies (M$\sim$10$^{10}\Msun$) can 
have 10-20\% of their masses driven down to several 100 pc within $\leq$10$^{8}$yr by an 
interaction. While the tidal triggering of disk instabilities with subsequent 
gas inflow is a short process, the bar formation leads to a continuous, low level nuclear 
feeding for much longer periods. Seyferts at these stages might appear to 
reside in normal environments or be ``isolated''. The life-time of this activity cycle (that is the
time needed for an AGN to exhaust its fuel) is comparable to the dynamical time scale of a galaxy
(of the order of 10$^{8}$ yr, depending on the black hole luminosity) and  
roughly that of the tidally induced bar of spiral ($\sim$10$^{9}$yr) assuming gas inflow rates
of $\sim$0.5$\Msun$yr$^{-1}$ (\eg \cite{byrd87,noguchi88,turner91,osterbrock93}). A latency period (before material enters 
the nuclear regions) of $\sim$2$\times$10$^{8}$ yr (depending on the galaxy mass) precedes the bar 
phase.

Several additional factors may modify this scenario.
A strong {\em and} centrally concentrated bulge would stabilize the disk against bar formation 
and thus prevent gas inflow until the final merging of the galaxies. Moreover, prograde 
encounters are more likely to trigger gas inflow through a bar instability compared to retrograde
encounters (in fact the generally less spectacular tidal tails in IR-selected interacting 
galaxies compared to their optical counterparts, suggest a broad range of spin orientations as 
opposed to the prograde-prograde encounters in most optical pairs). Consequently, these factors will 
also affect the level and life-time of nuclear activity.

Once tidal distortions of the galactic disk have occurred and circumnuclear star 
formation is triggered, material loses angular momentum
and falls further inwards. This has the effect of both feeding the AGN and obscuring the
broad line region: the galaxy appears as a Seyfert 2 type. At this time, the 
contribution of the circumnuclear starburst to the luminosity output of the system might be 
larger than that of the hidden AGN. Later, when dusty material is consumed in forming stars or 
is blown away by stellar winds we start seeing the BLR and thus we classify the galaxy as 
Seyfert 1. The time interval to progress from Seyfert 2 to Seyfert 1 may coincide with the time 
interval for a small companion to merge via a double nucleus phase into an ``isolated'' host showing 
tidal appendages. The double nucleus systems are in fact 
short-lived and thus relatively rare phases to be observed. The estimated time before they 
coalesce into a single remnant is relatively short (a few rotational periods) and their inner
regions will be relaxed in only $\sim$10$^{8}$ yr, which is comparable (or shorter) to the 
Seyfert activity lifetime. Given the tendency for Seyfert 1s to reside in early type hosts and
our finding of a large fraction of them to reside in isolated systems showing tidal appendages
emanating from a single nucleus, we suggest that these might be indeed recent merger products.

The phase corresponding to strong circumnuclear star formation triggered by the encounter
would probably show up through warm 60-100 $\mu$m and colder 25-60 $\mu$m colours, within the
scenario that we outlined in the previous section. This is not a well determined stage either,
the star formation physics seems to be playing an important role in the strength and duration of 
the starburst phase. In fact, star formation and feedback seems to be one of the outstanding 
problems in trying to model the dynamical evolution of these systems. Indeed it was recently 
shown through case studies of ULIRGs that they can be in very different dynamical stages,
although all associated with late stage mergers; this indicates that there is no unique 
trigger for ULIRG activity. In any case, as \cite{hutchings91} notes, since there exist IR-weak
AGNs, the IR activity cycle must be shorter than the activity duty cycle; furthermore the distinct
location of the various galaxy groups in their ``IR-diagram'' indicates that the IR-luminosity decay
must be happening very fast.
(For the above outline we used several bibliographical references; the most relevant are
\cite{byrd86,byrd87,hernquist89,maiolino95,hibbard96,mihos96,barnes98,gerritsen98,mihos98,mihos99,hacyan99,yun99}.)
In summary, we have seen that not only the gas content of the galaxy but also its internal 
structure, the orbital geometry of the encounter and the starburst physics are all important 
factors for determining the timing, duration and {\em level} of nuclear activity.

\begin{figure}
\epsscale{1.}
\plotfiddle{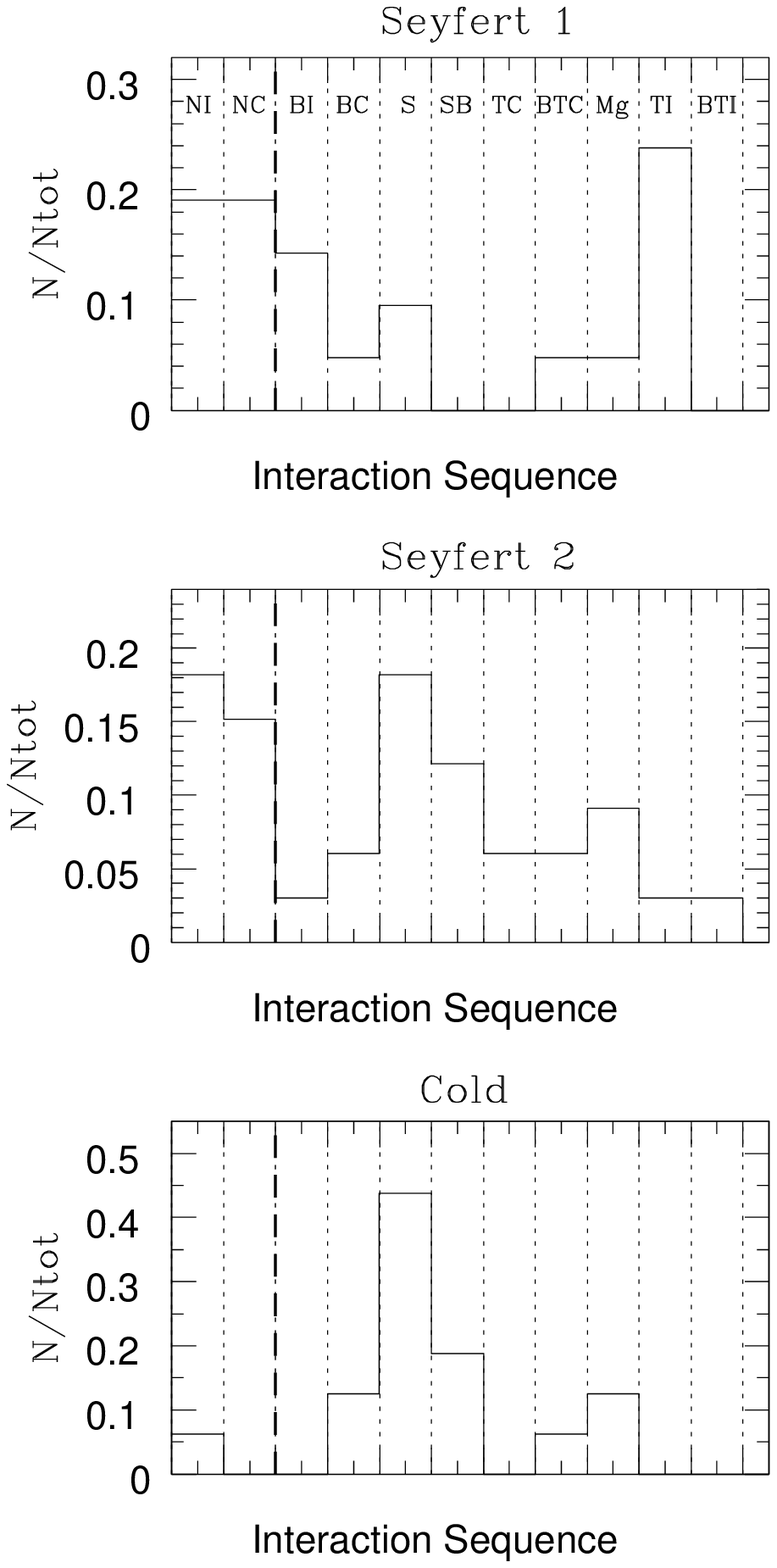}{400pt}{0}{50}{80}{-160}{-130}
\caption{Distribution of the three (sub)samples along the interaction sequence defined in the text (code: NI=normal isolated, NC=normal companions, BI=bar/ring isolated, BC=bar/ring companions, S=strongly interacting, SB=strongly interacting bar/ring, TC=tidal companions, BTC=bar/ring tidal companions, Mg=merger, TI=tidal isolated, BTI=bar/ring tidal isolated). \label{f6}}
\end{figure}

\placefigure{f6}
   
Based on the ideas described above we can now combine our earlier classification in terms of 
morphologies and interaction strength in a common sequence and plot the distribution of our
three (sub)samples along this sequence (Figure~\ref{f6}). This is by no means
 a unique description of the interaction process, given the 
several factors that affect the dynamical evolution of an encounter. However it is instructive 
for our ideas about the possible origins of the nuclear and IR activity to visualize the 
distribution of our samples through such a sequence. 
The logic we followed is simple: 

Objects with normal morphologies, either isolated (NI) or with companion galaxies (NC) but no 
morphological indication of interactions are put together in the beginning of this sequence. 
These might be objects that suffered a past encounter either too weak or too long ago to affect 
significantly their morphologies, but enough to trigger nuclear activity. 

Then, the purely barred/ringed systems in isolated (BI) or systems with companions (BC)
are grouped together. It is not clear where in the time evolution of the encounter these cases
belong. They might be representing low level interactions because of distant companions
or/and subsequent phases after perigalacticon passage, bar formation and nuclear activation. 

Right after, we have the strongly interacting systems, either
with (SB) or without (S) a bar/ring formation. These must be representing the phases at or 
shortly after perigalacticon, where starburst and subsequent nuclear activation occur. 

As these systems evolve in time, the companion separation increases but they show bars/rings and
strong tidal features (TC or BTC). 

The final stage we try to depict is the coalescence stage where either the two nuclei are still 
visible within a common body (Mg) or only a tidal appendage emanates from a single nucleus, 
isolated galaxy (TI). 

For the case denoted (BTI), an isolated object with a bar structure and a tidal feature, it is 
not clear whether it belongs here or to some other stage. We have observed only one Seyfert 2 in
this stage and perhaps it belongs to the (BTC) phase but the companion galaxy is too faint or is
farther than our search radius.
  
As expected from the earlier discussion of our results, the three (sub)samples show markedly 
different distributions (Figure~\ref{f6}): 

(i) The Seyfert 1 galaxies show a bimodal distribution peaking at normal morphologies
and advanced mergers, that is, the late stages of the interaction process according to our 
definition; there is a remarkable avoidance of the peak interaction phases or the ones right after
perigalacticon. There are however a few observed cases of strongly interacting Seyfert 1s.
How are we supposed to understand them? As we already pointed, the evolutionary
sequence we claim between the Seyfert 1 and 2 types is not one-fold (time) but depends as well
on the physical properties (dust content) of the host galaxy and maybe also on the interaction
parameters (how much material and how fast is driven inwards). 
 
(ii) The strong interaction phase represents precisely the peak in the distribution of 
Seyfert 2s in Figure~\ref{f6}, followed by a slow fall-off in particular towards the 
double nucleus mergers. The least probable occurrence of Seyfert 2s is in the isolated plus tidal
or bar/ring phases, because we think at those stages the galaxy is observed as Seyfert 1. 

(iii) 33\% of the Seyfert 2s occur in the sequence with normal morphologies (NI+NC); this is similar
to the Seyfert 1 fraction in these classes (38\%). These reside in hosts with a wide range of
environments. They have
comparable $L_{25}$ and in general a similar (large) range in optical and IR luminosities.
We have also mentioned throughout this paper the bimodality in some of the properties of Seyfert 
2s. We are thus lead to the conclusion that these might indeed be ``unified'' with Seyfert 1s 
within the context of the orientation unification model. 

(iv) The Cold sample objects show a narrow distribution in Figure~\ref{f6}, confined within the 
disruptive phases of the encounters. It is actually remarkable that after excluding the Seyfert 2s 
with normal morphologies (which we ``unified'' with Seyfert 1s) the Cold and Warm Seyfert 2 
distributions are very similar.

{\em We are thus led to the conclusion that this diagram represents primarily an evolutionary sequence
from bottom to top}.

\section{Conclusions}

We confirm that nuclear activity is linked to galactic interactions.

We find substantial evidence for an evolutionary sequence between the mid-IR Warm and Cold 
samples and between the two Seyfert types:

1. Warm Seyfert types 1 and 2 differ in two fundamental ways: time evolution and nuclear 
obscuration. Strongly interacting Seyfert 2s are in an earlier interaction stage than Seyfert 
1s. Seyfert 2s with normal morphologies can be unified with the type 1s in the context of 
orientation/obscuration models. 

2. Cold IR-selected galaxies are predominantly disrupted systems and bear significant 
similarities with the (weaker) Seyfert 2 interactors. 

3. The development or absence of a mid-IR excess is a three-fold effect: (i) evolutionary
(ii) dust/gas content in the progenitors and (iii) interaction parameters.

\acknowledgments
I am grateful to my thesis advisors George Miley and Walter Jaffe for providing
me with stimulation and support throughout the completion of this project.
This research has made use of the NASA/IPAC Extragalactic Database (NED) 
which is operated by the Jet Propulsion Laboratory, California Institute of 
Technology, under contract with the National Aeronautics and Space 
Administration. Part of this work was completed while the author held a 
National Research Council - NASA GSFC Research Associateship.

%
%

\begin{deluxetable}{lrrrrrrrrrrrrrrr}
\tablecolumns{16}
\tablewidth{0pc}
\tablecaption{Morphological and Interaction Classification. \label{tab1}}
\tablehead{
\colhead{} & \multicolumn{3}{c}{\itshape Normal} & \multicolumn{3}{c}{\itshape B/R} & \multicolumn{3}{c}{\itshape Tidal} & \multicolumn{3}{c}{\itshape Tidal+B/R} & \multicolumn{3}{c}{\itshape Mergers} \\
\cline{1-16}
\colhead{} & \colhead{S1} & \colhead{S2} & \colhead{C} & \colhead{S1} & \colhead{S2} & \colhead{C} & \colhead{S1} & \colhead{S2} & \colhead{C} & \colhead{S1} & \colhead{S2} & \colhead{C} & \colhead{S1} & \colhead{S2} & \colhead{C} \\
}
\startdata
{\itshape Isolated} & 19\% & 18\% & 6\% & 14\% & 3\% & \nodata & 24\% & 3\% & \nodata & \nodata & 3\% & \nodata & \nodata & \nodata & \nodata \nl
{\itshape Companions} & 19\% & 15\% & \nodata & 5\% & 6\% & 12.5\% & \nodata & 6\% & \nodata & 5\% & 6\% & 6\% & \nodata & \nodata & \nodata \nl
{\itshape Interacting} & 9.5\% & 18\% & 44\% & \nodata & 12\% & 19\% & \nodata & \nodata & \nodata & \nodata & \nodata & \nodata & \nodata & \nodata & \nodata \nl
{\itshape Mergers} & \nodata & \nodata & \nodata & \nodata & \nodata & \nodata & \nodata & \nodata & \nodata & \nodata & \nodata & \nodata & 5\% & 9\% & 12.5\% \nl
\enddata
\tablecomments{Distribution of the Warm Seyfert 1 and 2 and the Cold samples in Morphological and Interaction Classes, as defined in the text.}
\end{deluxetable}

\clearpage

\begin{deluxetable}{lrrrrrrrr}
\tablecolumns{9} 
\tablewidth{0pc}
\tablecaption{Median values and Standard Deviations for derived Quantities. \label{tab2}}
\tablehead{
\colhead{} & \multicolumn{4}{c}{\itshape Brightest Companions} & \multicolumn{4}{c}{\itshape Closest Companions} \\
\cline{1-9}
\colhead{} & \colhead{$S/D_{p}$} & \colhead{$log[Q_{D}]$} & \colhead{$D_{s}/D_{p}$} & \colhead{$M_{s}-M_{p}$} & \colhead{$S/D_{p}$} & \colhead{$log[Q_{D}]$} & \colhead{$D_{s}/D_{p}$} & \colhead{$M_{s}-M_{p}$} \\
}
\startdata
\cutinhead{{\itshape Seyfert 1}}
 {\em Median} & 1.57 & -0.91 & 0.8 & 1.74 & 1.41 & -0.82 & 0.70 & 1.56 \nl
 $\sigma$ & 1.55 & 1.25 & 0.50 & 1.51 & 1.19 & 1.11 & 0.50 & 1.47 \nl
\cutinhead{{\itshape Seyfert 2}}
 {\em Median} & 1.20 & -0.91 & 0.50 & 1.59 & 0.88 & -0.54 & 0.50 & 1.88 \nl
 $\sigma$ & 1.09 & 1.20 & 0.51 & 1.70 & 1.09 & 1.16 & 0.49 & 1.61 \nl
\cutinhead{{\itshape Cold}}
 {\em Median} & 0.56 & 0.16 & 0.60 & 1.79 & 0.78 & -0.07 & 0.70 & 1.61 \nl
 $\sigma$  & 0.81 & 1.02 & 0.30 & 1.21 & 0.98 & 1.19 & 0.32 & 1.24 \nl
\enddata
\tablecomments{$S/D_{p}$: Separation of two galaxies normalized to the major axis diameter of the primary.
$Q_{D}$: Interaction strength (as defined in text).
$D_{s}/D_{p}$: Ratio of major axis diameters of companion and primary galaxies.
$M_{s}-M_{p}$: Optical magnitude (mostly $R$) difference between companion and primary.}
\end{deluxetable}

\begin{deluxetable}{lrrrrrr}
\tablecolumns{7}
\tablewidth{0pc}
\tablecaption{Merger Properties. \label{tab3}}
\tablehead{
\colhead{} & \multicolumn{4}{c}{\itshape Warm} & \multicolumn{2}{c}{\itshape Cold} \\
\cline{1-7}
\colhead{} & \colhead{S1\tablenotemark{1}} & \colhead{S2\tablenotemark{2}} & \colhead{S2\tablenotemark{3}} & \colhead{S2\tablenotemark{4}} & \colhead{-\tablenotemark{5}} & \colhead{S2\tablenotemark{6}} \\ 
}
\startdata
$S_{nuclear}$ & 2 & 4 & 9 & \nodata & \nodata & 3.3 \nl
$|\Delta m|_{nuc}$ & 0.61 & 0.56 & 0.64 & \nodata & \nodata & \nodata \nl
$(B-V)_{nuc}$ & 0.85/0.59 & 1.45/0.55 & 0.99/0.80 & \nodata & \nodata & \nodata \nl
$(B-R)_{nuc}$ & 1.29/0.88 & 1.59/1.05 & 1.92/1.47 & \nodata & \nodata & \nodata \nl  
$L_{B}[disk]$ & 6.29 & 7.08 & 6.86 & \nodata & 6.64 & \nodata \nl
$L_{R}[disk]$ & 5.94 & 6.81 & 6.58 & \nodata & 6.19 & \nodata \nl
$L_{25}$ & 10.59 & 11.22 & 11.36 & 11.44 & 10.27 & 11.12 \nl
$L_{FIR}$ & 10.64 & 11.28 & 11.84 & 11.79 & 11.26 & 11.88 \nl
$\alpha_{(25,60)}$ & -0.33 & -0.35 & -1.46 & -1.09 & -2.80 & -2.19 \nl
$\alpha_{(60,100)}$ & -0.54 & +0.30 & -0.30 & +0.37 & -0.47 & -0.03 \nl
\enddata
\tablecomments{Properties tabulated as a function of nuclear type for the Warm and Cold sample mergers. Row 1: nuclear separation in kpc, Row 2: absolute R magnitude difference between the two nuclei, Rows 3-4: optical colours for the two nuclei, Rows 5-6: logarithm of optical luminosities in solar units $\Lsun$, Rows 7-8: logarithm IR luminosities in solar units $\Lsun$, Rows 9-10: IR colours.}
\tablecomments{Listed objects: S1$^{1}$: IRAS 19580-1818, S2$^{2}$: IRAS 13536+1836, S2$^{3}$: IRAS 19254-7245, S2$^{4}$: IRAS 00198-7926, -$^{5}$: IRAS 03531-4507, S2$^{6}$: IRAS 23128-5919. }
\end{deluxetable}

%
%

\clearpage

\clearpage

\clearpage

\figcaption[f5.ps]{$R$-band grey-scale images and overplotted contours for the six mergers in our Warm and Cold samples. The image sizes are: IRAS 1958--1818: 19.8$\times$19.8 kpc, IRAS 13536+1836: 130$\times$130 kpc, IRAS 19254-7245: 270$\times$270 kpc, IRAS 00198-7926: 72$\times$72 kpc, IRAS 23128-5919: 83$\times$83 kpc and IRAS 03531-4507: 180.4$\times$180.4 kpc. \label{f5}}

\end{document}